\documentclass{article}

\usepackage[margin=1.5in,nohead]{geometry}
\usepackage{amsmath}	
\usepackage{amsfonts}
\usepackage{amssymb}    
\usepackage{setspace}                                           
\usepackage{graphicx}
\usepackage{algorithm, algpseudocode}
\usepackage{color,soul}
\usepackage{authblk}
\providecommand{\U}[1]{\protect\rule{.1in}{.1in}}

\newtheorem{assumption}{Assumption}

\newtheorem{definition}{Definition}

\newtheorem{notation}{Notation}
\newtheorem{problem}{Problem}
\newtheorem{proposition}{Proposition}
\newtheorem{remark}{Remark}

\newcommand{\tabincell}[2]{\begin{tabular}{@{}#1@{}}#2\end{tabular}}
\let\oldIEEEkeywords\IEEEkeywords
\def\IEEEkeywords{\oldIEEEkeywords\normalfont\bfseries\ignorespaces}
\author{Zhe~Xu, Melkior Ornik, A. Agung~Julius, Ufuk Topcu
	\thanks{Zhe~Xu and Melkior Ornik are with the Institute
		for Computational Engineering and Sciences (ICES), University of Texas,
		Austin, Austin, TX 78712, A. Agung~Julius is with the Department of Electrical, Computer, and Systems Engineering, Rensselaer Polytechnic Institute, Troy, NY 12180, Ufuk Topcu is with the Department
		of Aerospace Engineering and Engineering Mechanics, and the Institute
		for Computational Engineering and Sciences (ICES), University of Texas,
		Austin, Austin, TX 78712, e-mail: zhexu@utexas.edu, mornik@ices.utexas.edu, juliua2@rpi.edu, utopcu@utexas.edu.}
}

\begin{document}
	
\title{Information-Guided Temporal Logic Inference with Prior Knowledge}

\maketitle

\begin{abstract}               
 This paper investigates the problem of inferring knowledge from data so that the inferred knowledge is interpretable and informative to humans who have prior knowledge. Given a dataset as a collection of system trajectories, we infer parametric linear temporal logic (pLTL) formulas that are informative and satisfied by the trajectories in the dataset with high probability. The informativeness of the inferred formula is measured by the information gain with respect to given prior knowledge represented by a prior probability distribution. We first present two algorithms to compute the information gain with a focus on two types of prior probability distributions: stationary probability distributions and probability distributions expressed by discrete time Markov chains. Then we provide a method to solve the inference problem for a subset of pLTL formulas with polynomial time complexity with respect to the number of Boolean connectives in the formula. We provide implementations of the proposed approach on explaining anomalous patterns, patterns changes and explaining the policies of Markov decision processes.
\end{abstract}                                                                       

\section{Introduction}
\begin{figure}[th]
	\centering
	\includegraphics[width=10cm]{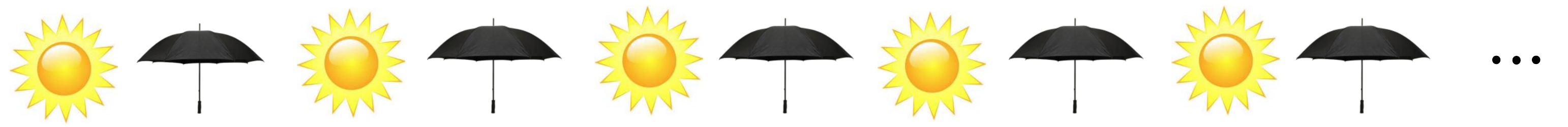}\caption{The dataset of weather conditions on consecutive days in a certain geographical region.}
	\label{weatherData}
\end{figure}

\begin{figure}[th]
	\centering
	\includegraphics[width=6cm]{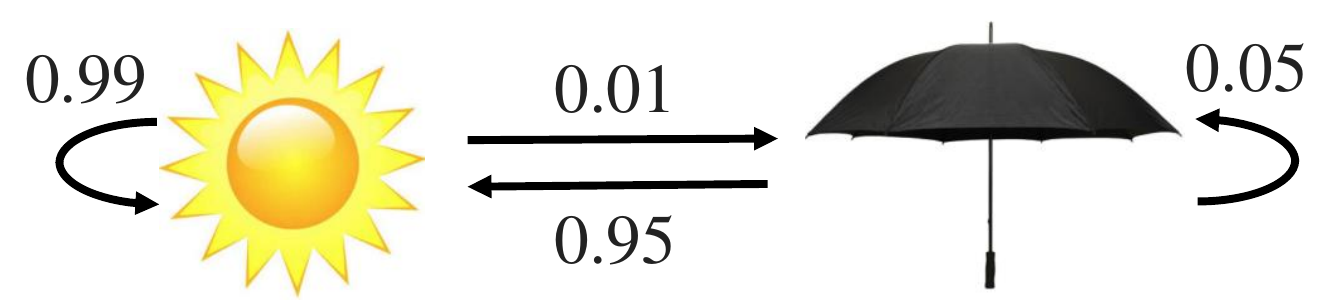}\caption{The prior knowledge represented as a discrete time Markov chain (with two simplified weather conditions: sunny and rainy).}
	\label{weatherMC}
\end{figure}

Inferring human-interpretable knowledge from execution trajectories of a system is important in many applications. Such knowledge, for example, may represent behaviors that deviate from the expected behaviors according to some prior knowledge. These deviations may unveil some unknown patterns of the system, or indicate that changes or anomalies have occurred in the system. As a simple example, suppose that we are given a recently collected dataset of weather conditions on consecutive days in a certain geographical region, as shown in Fig. \ref{weatherData}. Furthermore, suppose that we also have prior knowledge on historical weather conditions in the region, represented as a discrete time Markov chain as shown in Fig. \ref{weatherMC}. We raise the following question: What knowledge can we infer from the dataset so that it is \textit{interpretable} and \textit{informative} to humans who have the possibly outdated impression (i.e. prior knowledge)?                   

The interpretability of the inferred knowledge represents the extent that humans can understand the knowledge. In the past decade, there has been a growing interest in inferring temporal logic formulas from system trajectories \cite{Asarin2012, Fainekos2012,Jin13,zhe2016}. The temporal logic formulas can express features and patterns in a form that resembles natural languages \cite{Kong2017} and thus improve the interpretability of the
inferred knowledge.

The informativeness represents the extent to which the inferred knowledge deviates from prior knowledge. In the example on weather conditions, suppose that we are given two candidate temporal logic formulas: one reads as ``whenever it is rainy, the next day will be sunny'' and the other one reads as ``whenever it is sunny, the next day will be rainy''. While both formulas are consistent with the data as shown in Fig. \ref{weatherData}, clearly the second one is more informative in describing how the weather patterns have deviated from the prior knowledge as shown in Fig. \ref{weatherMC}.

There are various methods for inferring temporal logic formulas from data. For example, the authors in \cite{Kong2017} developed an inference method to automatically select both the structure and the parameters of signal temporal logic (STL) formulas that classify the trajectories between a desired set and an undesired set. In \cite{Bombara2016}, the authors proposed a decision-tree approach to infer STL formulas for classification. In \cite{Neider}, the authors presented methods to infer linear temporal logic formulas from data through a reduction to a series of satisfiability problems in
propositional Boolean logic. However, to the best of our knowledge, none of the existing approaches performs the inference task while considering the informativeness of the inferred formulas over prior knowledge. 

In this paper, we propose a theoretical framework and algorithms for the information-guided temporal logic inference. We infer a parametric linear temporal logic formula that is consistent with a set of trajectories and provides a high information gain (the notion of information gain will be formalized in Section \ref{Sec_problem}) over a given prior probability distribution. We conduct three case studies as implementations of the proposed approach in the following applications.
\begin{itemize}
\item \textit{Explaining anomalous patterns and pattern changes}:                                                                
If we assume that normal behavior occur with high probability based on the prior probability distribution, then the inferred formulas can be used to explain anomalous patterns ``hidden'' in the dataset. If the prior probability distribution represents outdated knowledge or impressions, then the inferred formulas can be used to explain pattern changes at present. 
\item \textit{Explaining policies of Markov decision processes}:
If the dataset consists of observed trajectories of a Markov decision process (MDP), then the inferred temporal logic formulas may be used as explanations of the policies of the MDP. 
\end{itemize}

\section{Parametric Linear Temporal Logic} 
In this section, we present an overview of parametric linear temporal logic
(pLTL) \cite{pLTL2014,pLTL}. We start with the syntax and semantics of pLTL. The domain $\mathbb{B}=\{\top, \bot\}$ ($\top$ and $\bot$ represents True and False respectively) is
the Boolean domain and the time index set is a discrete set of natural numbers $\mathbb{T}=\{1, 2, \dots\}$. We assume that there is an underlying system $\mathcal{H}$. The state $s$ of the
system $\mathcal{H}$ belongs to a finite set of states $S$. A trajectory $s_{1:L}=s_1s_2\cdots s_L$ describing an evolution of the system $\mathcal{H}$ is a function from $\mathbb{T}$ to                    
$S$. A set $\mathcal{AP}=\{\pi_{1},\pi_{2},\dots,\pi_{n}\}$ is a set of atomic predicates. $\mathcal{L}: S\rightarrow2^{\mathcal{AP}}$ is a labeling function assigning a subset of atomic predicates in $\mathcal{AP}$ to each state $s\in S$.                                                                  

The syntax of the pLTL formula is defined recursively as follows:
\[
\begin{split}
\phi:=&\top\mid\pi\mid\lnot\phi\mid\phi_{1}\wedge\phi_{2}\mid \bigcirc\phi\mid \phi_1\mathcal{U}\phi_2 \mid \Diamond_{\sim i}\phi,
\label{syntax}
\end{split}
\]
where $\pi$ is an atomic
predicate, $\lnot$ and $\wedge$ stand for negation and conjunction respectively, $\bigcirc$ and $\mathcal{U}$ are temporal operators representing \textquotedblleft next\textquotedblright~and \textquotedblleft until\textquotedblright~respectively, $\Diamond_{\sim i}$ $(i\in\mathbb{T})$ is a parametrized temporal operator representing \textquotedblleft parametrized eventually\textquotedblright, where $\sim\in\{\ge,\le\}$, $i$ is a temporal parameter. We can also derive $\vee$ (disjunction), $\Diamond$ (eventually), $\Box$ (always), $\mathcal{R}$ (release), $\Box_{\sim i}$ (parametrized always), $\mathcal{U}_{\sim i}$ (parametrized until), $\mathcal{R}_{\sim i}$ (parametrized release) from the above-mentioned operators as described in \cite{pLTL2014}. We can also derive $\Diamond_{\ge i_1, \le i_2}$ and $\Box_{\ge i_1, \le i_2}$ ($i_1<i_2$) as
\[
\begin{split}
\Diamond_{\ge i_1, \le i_2}\phi=\Diamond_{\ge i_1}\phi\wedge\Diamond_{\le i_2}\phi,\\
\Box_{\ge i_1, \le i_2}\phi=\Box_{\ge i_1}\phi\wedge\Box_{\le i_2}\phi.
\end{split}
\] 

Next, we introduce the Boolean semantics of a pLTL formula in the strong and the weak view \cite{Ho2014}, which is used in evaluating the satisfaction or violation of pLTL formula by trajectories of finite length. 

The satisfaction relation $(s_{1:L},k)\models_{\rm{S}}\phi$ for trajectory $s_{1:L}$ at time index $k$ with respect to a pLTL formula $\phi$ as Boolean semantics in the strong view is defined recursively as follows: 
\[
\begin{split}
(s_{1:L},k)\models_{\rm{S}}\pi\quad\mbox{iff}\quad& k\le L~\mbox{and}~\pi\in\mathcal{L}(s_k),\\
(s_{1:L},k)\models_{\rm{S}}\lnot\phi\quad\mbox{iff}\quad & (s_{1:L},k)\not\models_{\rm{\rm{W}}}\phi,\\
(s_{1:L},k)\models_{\rm{S}}\phi_{1}\wedge\phi_{2}\quad\mbox{iff}\quad & (s_{1:L},k)\models_{\rm{S}}\phi_{1}\quad\\& \mbox{and}\quad(s_{1:L},k)\models_{\rm{S}}\phi_{2}\\
(s_{1:L},k)\models_{\rm{S}}\bigcirc\phi\quad\mbox{iff}\quad& (s_{1:L},k+1)\models_{\rm{S}}\phi,\\
 \end{split}
\] 
\[
\begin{split}
(s_{1:L},k)\models_{\rm{S}}\phi_{1}\mathcal{U}\phi_{2}\quad\mbox{iff}\quad &  \exists
 k'\ge k, s.t.(s_{1:L},k')\models_{\rm{S}}\phi_{2},\\
  &  (s_{1:L},k^{\prime\prime})\models_{\rm{S}}\phi_{1}, \forall k^{\prime\prime}\in[k, k']\\
(s_{1:L},k)\models_{\rm{S}}\Diamond_{\sim i}\phi\quad\mbox{iff}\quad & \exists
k'\sim k+i, \mbox{s.t.}~(s_{1:L},k')\models_{\rm{S}}\phi.
\end{split}
\]

The satisfaction relation $(s_{1:L},k)\models_{\rm{W}}\phi$ as Boolean semantics in the weak view is defined recursively as follows:
\[
\begin{split}
(s_{1:L},k)\models_{\rm{W}}\pi\quad\mbox{iff}\quad\mbox{iff}\quad& \textrm{either of the following holds}:\\
	& 1)~k>L;~2)~k\le L~\mbox{and}~\pi\in\mathcal{L}(s_k),
	 \\	
(s_{1:L},k)\models_{\rm{W}}\lnot\phi\quad\mbox{iff}\quad & (s_{1:L},k)\not\models_{\rm{\rm{S}}}\phi,
\end{split}
\] 
while the Boolean semantics of the other logical and temporal operators can be obtained from those in the strong view by replacing each $\models_{\rm{S}}$ with $\models_{\rm{W}}$.  

If the satisfaction relations are evaluated at time index $k=1$, then we write $s_{1:L}\models_{\rm{S}}\phi$ or $s_{1:L}\models_{\rm{W}}\phi$ for brevity.                

Intuitively, if a trajectory of finite length can be extended to infinite length, then the strong view indicates that the truth value of the formula on the infinite-length trajectory is already ``determined'' on the trajectory of finite length, while the weak view indicates that it may not be ``determined'' yet \cite{Ho2014}. As an example, a trajectory $s_{1:3}=s_1s_2s_3$ is not possible to strongly satisfy $\phi=\Box_{\le5}\pi$, but $s_{1:3}$ is possible to strongly violate $\phi$, i.e., $(s_{1:L},k)\models_{\rm{S}}\lnot\phi$ is possible.

Trajectories of finite length are sufficient to strongly satisfy (resp. violate) \textit{syntactically co-safe} (resp. \textit{safe}) pLTL formulas \cite{KupfermanVardi2001}, which are defined in the following definitions.

\begin{definition}
	The syntax of the \textit{syntactically co-safe} pLTL formula is defined recursively as follows:
	\[
	\begin{split}
	\phi:=&\top\mid\pi\mid\lnot\pi\mid\phi_{1}\wedge\phi_{2}\mid\phi_{1}\vee
	\phi_{2}\mid \bigcirc\phi\mid \Diamond\phi\mid \phi_1\mathcal{U}\phi_2\\& \mid \Diamond_{\sim i}\phi\mid \Box_{\le i}\phi\mid \phi_1\mathcal{U}_{\sim i}\phi_2\mid \phi_1\mathcal{R}_{\le i}\phi_2.
	\end{split}
	\]
\end{definition}  

\begin{definition}
	The syntax of the \textit{syntactically safe} pLTL formula is defined as follows:
	\[
	\begin{split}
	\phi:=&\bot\mid\pi\mid\lnot\pi\mid\phi_{1}\wedge\phi_{2}\mid\phi_{1}\vee
	\phi_{2}\mid \bigcirc\phi\mid \Box\phi\mid \phi_1\mathcal{R}\phi_2\\& \mid \Diamond_{\le i}\phi\mid \Box_{\sim i}\phi\mid \phi_1\mathcal{U}_{\le i}\phi_2\mid \phi_1\mathcal{R}_{\sim i}\phi_2.
	\end{split}
	\]
\end{definition}

There are pLTL formulas that are neither syntactically co-safe nor syntactically safe pLTL formulas, such as $\Box\Diamond_{\ge i} \pi$. These pLTL formulas can neither be strongly satisfied nor strongly violated by trajectories of finite length. As we intend to infer pLTL formulas from trajectories of finite length (as in a dataset), in this paper we only focus on syntactically co-safe or syntactically safe pLTL formulas, which will be referred to as (co-)safe pLTL formulas for brevity.  

In the following, for simplicity we require that the trajectory of finite length is sufficiently long so that it can strongly satisfy (resp. violate) the corresponding syntactically co-safe (resp. safe) formula. With slight abuse of notation, we use the notations $s_{1:L}\models\phi$ and $s_{1:L}\not\models\phi$ to denote $s_{1:L}\models_{\rm{S}}\phi$ and $s_{1:L}\not\models_{\rm{S}}\phi$ if $\phi$ is syntactically co-safe, and to denote $s_{1:L}\models_{\rm{W}}\phi$ and $s_{1:L}\not\models_{\rm{W}}\phi$ if $\phi$ is syntactically safe. Additionally, the word ``satisfy'' without the modifying adverb ``strongly'' or ``weakly'' is meant in the strong (resp. weak) view if the corresponding formula $\phi$ is syntactically co-safe (resp. safe). If a pLTL formula $\phi$ is both syntactically co-safe and syntactically safe, then the strong view and the weak view are equivalent for $\phi$ as we require that the trajectory of finite length is sufficiently long so that it can strongly satisfy and strongly violate the formula.                       

The size of a (co-)safe pLTL formula $\phi$, denoted as $\ell(\phi)$, is defined as the number of Boolean connectives in $\phi$. Note that logically equivalent formulas may have different sizes.

\section{Problem Formulation}
\label{Sec_problem}
We denote $\mathcal{B}_L$ as the set of all possible trajectories with length $L$ of the underlying system $\mathcal{H}$. We are also given a dataset $\mathcal{S}_L=\{\hat{s}^1_{1:L}, \dots, \hat{s}^m_{1:L}\}\subset\mathcal{B}_L$ as a collection of trajectories and a prior probability distribution over $\mathcal{B}_L$. 

\begin{notation}
We use $\mathcal{F}_L: \mathcal{B}_L\rightarrow[0,1]$ to denote the prior probability distribution, $\mathbb{P}_{\mathcal{B}_L,\phi}$ to denote the probability of a trajectory $s_{1:L}$ satisfying $\phi$ in the set $\mathcal{B}_L$ based on $\mathcal{F}_L$, and $\mathbb{\bar{P}}_{\mathcal{S}_L,\phi}:=\vert\{\hat{s}_{1:L}: \hat{s}_{1:L}\in\mathcal{S}_L, \hat{s}_{1:L}\models\phi\}\vert/\vert\mathcal{S}_L\vert$ to denote the empirical probability of a trajectory $\hat{s}_{1:L}$ satisfying $\phi$ in the dataset $\mathcal{S}_L$ (i.e. the ratio of the number of trajectories in $\mathcal{S}_L$ that satisfy $\phi$), where $\vert S \vert$ denotes the cardinality of a set $S$.
\end{notation}

\begin{assumption}
	We assume that every trajectory in $\mathcal{B}_L$ occurs with non-zero probability according to $\mathcal{F}_L$. 
	\label{assume}
\end{assumption}

\begin{definition}
Given a prior probability distribution $\mathcal{F}_L$ that satisfies Assumption \ref{assume}, we define $\mathcal{\bar{F}}^{\phi}_L: \mathcal{B}_L\rightarrow[0,1]$ as the estimated posterior probability distribution given the pLTL formula $\phi$ and the dataset $\mathcal{S}_L$, which is expressed as
\[                           
\begin{split}
\mathcal{\bar{F}}^{\phi}_L(s_{1:L})=
\begin{cases}
\frac{\mathcal{F}_L(s_{1:L})\mathbb{\bar{P}}_{\mathcal{S}_L,\phi}}{\mathbb{P}_{\mathcal{B}_L,\phi}} ~~~~~~~~~~~\mbox{if}~s_{1:L}\models\phi,\\ 
\frac{\mathcal{F}_L(s_{1:L})\mathbb{\bar{P}}_{\mathcal{S}_L,\lnot\phi}}{\mathbb{P}_{\mathcal{B}_L,\lnot\phi}}~~~~~~~~~~\mbox{if}~s_{1:L}\not\models\phi.
\end{cases} 
\end{split}
\]
\label{know}
\end{definition}
 
\begin{remark}
	According to Assumption \ref{assume}, for any $s_{1:L}\in\mathcal{B}_L$, if $s_{1:L}\models\phi$, then $\mathbb{P}_{\mathcal{B}_L,\phi}>0$; if $s_{1:L}\not\models\phi$, then $\mathbb{P}_{\mathcal{B}_L,\lnot\phi}>0$. 
\end{remark} 

\begin{remark}
We require that 
\[                           
\begin{split}
\sum\limits_{s_{1:L}\in\mathcal{B}_L, s_{1:L}\models\phi}\mathcal{\bar{F}}^{\phi}_L(s_{1:L})=\mathbb{\bar{P}}_{\mathcal{S}_L,\phi},
\end{split}
\]
which means the formula $\phi$ is true with the same empirical probability in $\mathcal{S}_L$ as the probability in $\mathcal{B}_L$ based on $\mathcal{\bar{F}}^{\phi}_L$. Then the expression of $\mathcal{\bar{F}}^{\phi}_L$ can be derived using Bayes' theorem.
\end{remark} 

\begin{remark}
	If $\mathbb{P}_{\mathcal{B}_L,\phi}=\mathbb{\bar{P}}_{\mathcal{S}_L,\phi}$, then $\mathcal{\bar{F}}^{\phi}_L(s_{1:L})=\mathcal{F}_L(s_{1:L})$ for any $s_{1:L}\in\mathcal{B}_L$, which means that the estimated posterior probability distribution given $\phi$ and the dataset $\mathcal{S}_L$ is the same as the prior probability distribution if $\phi$ is true with the same empirical probability in $\mathcal{S}_L$ as the probability in $\mathcal{B}_L$ based on $\mathcal{F}_L$. Especially, if $\phi=\top$ (resp. $\phi=\bot$), then $\mathbb{P}_{\mathcal{B}_L,\phi}=\mathbb{\bar{P}}_{\mathcal{S}_L,\phi}=1$ (resp. $\mathbb{P}_{\mathcal{B}_L,\lnot\phi}=\mathbb{\bar{P}}_{\mathcal{S}_L,\lnot\phi}=1$), so $\mathcal{\bar{F}}^{\phi}_L(s_{1:L})=\mathcal{F}_L(s_{1:L})$ for any $s_{1:L}\in\mathcal{B}_L$, which means that the estimated posterior probability distribution given $\top$ ($\bot$) and any dataset $\mathcal{S}_L$ is the same as the prior probability distribution. 
\end{remark} 
                              
\begin{definition}
We define 
\[                           
\begin{split}
\mathcal{I}(\mathcal{F}_L,\mathcal{\bar{F}}^{\phi}_L):=\frac{1}{L}\cdot D_{\rm{KL}}(\mathcal{\bar{F}}^{\phi}_L\vert\vert\mathcal{F}_L)
\end{split}
\]
as the information gain when the prior probability distribution $\mathcal{F}_L$ is updated to the estimated posterior probability distribution $\mathcal{\bar{F}}^{\phi}_L$, where
\[                           
\begin{split}
D_{\rm{KL}}(\mathcal{\bar{F}}^{\phi}_L\vert\vert\mathcal{F}_L)=\sum\limits_{s_{1:L}\in\mathcal{B}_L}\mathcal{\bar{F}}^{\phi}_L(s_{1:L})\log\frac{\mathcal{\bar{F}}^{\phi}_L(s_{1:L})}{\mathcal{F}_L(s_{1:L})}
\end{split}
\] 
is the Kullback-Leibler divergence from $\mathcal{F}_L$ to $\mathcal{\bar{F}}^{\phi}_L$.
\label{KL}
\end{definition} 

\begin{problem} 
	Given a dataset $\mathcal{S}_L=\{\hat{s}^1_{1:L}, \dots,$ $\hat{s}^m_{1:L}\}$, a prior probability distribution $\mathcal{F}_L$, real constant $p_{\rm{th}}\in(0,1]$ and integer constant $\ell_{\rm{th}}\in(0, \infty)$, we construct a (co-) safe pLTL formula $\phi$ that maximizes the information gain $\mathcal{I}(\mathcal{F}_L,\mathcal{\bar{F}}^{\phi}_L)$ while satisfying the following two constraints:
	\begin{itemize}
		\item \textit{coverage constraint}: $\mathbb{\bar{P}}_{\mathcal{S}_L,\phi}\ge p_{\rm{th}}$, i.e.,
		the trajectories in $\mathcal{S}_L$ should satisfy $\phi$ with empirical probability at least $p_{\rm{th}}$;
		\item \textit{size constraint}:
		$\ell(\phi)\le\ell_{\rm{th}}$, i.e., the size of $\phi$ should not exceed $\ell_{\rm{th}}$. 
	\end{itemize}      
	\label{problem1}           
\end{problem}

Intuitively, the objective is to make the inferred pLTL formula informative over the prior probability distributions. The coverage constraint is to make the inferred pLTL formula $\phi$ consistent (with probability no less than $p_{\rm{th}}$) with the dataset. The size constraint is to make the inferred pLTL formula concise, as an unnecessarily long and complicated formula is too specific and interpretability is compromised. 

\section{Computation of Information Gain with Prior Probability Distributions}                  
\label{gain}
To solve Problem \ref{problem1}, one needs to compute the information gain for a (co-) safe pLTL formula. The following proposition relates $\mathcal{I}(\mathcal{F}_L,\mathcal{\bar{F}}^{\phi}_L)$ to the probabilities $\mathbb{P}_{\mathcal{B}_L,\phi}$ and $\mathbb{\bar{P}}_{\mathcal{S}_L,\phi}$.

\begin{proposition} 
	The information gain $\mathcal{I}(\mathcal{F}_L,\mathcal{\bar{F}}^{\phi}_L)$ satisfies
	\[                           
	\begin{split}
	\mathcal{I}(\mathcal{F}_L,\mathcal{\bar{F}}^{\phi}_L)=&\frac{1}{L}(\mathbb{\bar{P}}_{\mathcal{S}_L,\phi}\log(\mathbb{\bar{P}}_{\mathcal{S}_L,\phi})-\mathbb{\bar{P}}_{\mathcal{S}_L,\phi}\log(\mathbb{P}_{\mathcal{B}_L,\phi})\\&+(\mathbb{\bar{P}}_{\mathcal{S}_L,\lnot\phi})\log(\mathbb{\bar{P}}_{\mathcal{S}_L,\lnot\phi})-(\mathbb{\bar{P}}_{\mathcal{S}_L,\lnot\phi})\log(\mathbb{P}_{\mathcal{B}_L,\lnot\phi})).
	\end{split}
	\]	 
	\label{exp}   
\end{proposition} 

\textbf{proof}     
See Appendix.                                      
                                                
In the following two subsections, we present how to recursively compute the information gain through Proposition \ref{exp} for two specific types of prior probability distributions.

\subsection{Computation of Information Gain for Stationary Prior Probability Distributions}
\label{gain_A}
In this subsection, we present an algorithm to compute the information gain for stationary prior probability distributions.

We define a stationary prior probability distribution $\mathcal{F}_L: \mathcal{B}_L\rightarrow[0,1]$ as follows: for each trajectory $s_{1:L}\in\mathcal{B}_L$, the probability $\mathcal{F}_L(s_{1:L})=\mathcal{F}(s_{1})\mathcal{F}(s_{2})\dots\mathcal{F}(s_{L})$, where $\mathcal{F}$  is the probability distribution on $S$ which remains the same for any time index $k$.

\begin{definition}
	A deterministic finite automaton (DFA) is a tuple $\mathcal{A}=(\mathcal{Q},q_{0},\Sigma,\delta,Acc)$ where
	$\mathcal{Q}=\{q_0, q_1, \dots, q_K\}$ is a finite set of states, $q_{0}$ is the initial state, $\Sigma$ is the alphabet, $\delta:\mathcal{Q}\times\Sigma\rightarrow\mathcal{Q}$ is the transition relation, $Acc\subseteq2^{\mathcal{Q}}$ is a finite set of accepting states. 
	\label{sw}
\end{definition} 

For any syntactically co-safe (resp. safe) pLTL formula $\phi$, a DFA $\mathcal{A}^{\phi}=(\mathcal{Q},q_0,2^{\mathcal{AP}},\delta,Acc^{\phi})$ (resp. $\mathcal{A}^{\lnot\phi}=(\mathcal{Q},q_0,2^{\mathcal{AP}},\delta,Acc^{\lnot\phi})$) can be constructed with input alphabet $2^{\mathcal{AP}}$ that accepts all and only trajectories that strongly satisfy $\phi$ (resp. strongly violate $\phi$) \cite{KupfermanVardi2001}.

\begin{algorithm}
	\caption{Information gain for stationary prior probability distributions.}                                           
	\label{compute}
	\begin{algorithmic}[1]
		\State \textbf{Input:} $\mathcal{S}_L=\{\hat{s}^1_{1:L}, \dots, \hat{s}^m_{1:L}\}, \phi, \mathcal{F}_L$
		\State $\beta\gets\mathbb{\bar{P}}_{\mathcal{S}_L,\phi}$
		\State Obtain the DFA $\mathcal{A}^{\phi}$ (if $\phi$ is syntactically co-safe) or $\mathcal{A}^{\lnot\phi}$ (if $\phi$ is syntactically safe)
		\For{$k=0$ to $K$}  
		\State \textbf{if} $q_{k}\in Acc^{\phi}$ (resp. $Acc^{\lnot\phi}$) \textbf{then} $p(L,q_{k})\gets1$ 
		\State \textbf{else} $p(L,q_{k})\gets0$
		\State \textbf{end if}                              
		\EndFor		     
		\For{$\ell=L$ to 2, $j=0$ to $K$}                                                          
		\State $p(\ell-1,q_{j})\gets\sum_{k=0}^{K}c_{j,k}p(\ell,q_{k})$                                                                                    
		\EndFor
		\State \textbf{if} $\phi$ is syntactically co-safe \textbf{then} $\gamma\gets p(1,q_{0})$
		\State \textbf{else} $\gamma\gets 1-p(1,q_{0})$	
		\State \textbf{end if} 	
		\State return $\mathcal{I}=\frac{1}{L}(\beta\log(\frac{\beta}{\gamma})+(1-\beta)\log(\frac{1-\beta}{1-\gamma}))$	
	\end{algorithmic}
\end{algorithm}	

\begin{proposition} 
	For a (co-)safe pLTL formula $\phi$ and a stationary prior probability distribution $\mathcal{F}_L$, Algorithm \ref{compute} returns $\mathcal{I}(\mathcal{F}_L,\mathcal{\bar{F}}^{\phi}_L)$.
	\label{computeI1}
\end{proposition}  

\textbf{proof}     
See Appendix.  

\begin{remark}
	The time complexity of Algorithm \ref{compute} is $O(LK^2)$, where $K+1$ is the number of states of the DFA $\mathcal{A}^{\phi}$. For different (co-)safe pLTL formulas with the same predicates, same temporal operators but different temporal parameters, the computational cost can be reduced by storing the results of computation for one formula and reusing them in the computation for another formula. For example, for $\phi=\Box\Diamond_{\le i}\pi$, we have ($K=i+1$)\\
	\begin{align}
	\begin{split}
	\textbf{p}_L^{\phi}(\ell-1,\cdot)=
	\begin{bmatrix}
	\Pi_{i+1} & \mathbb{P}(\lnot\pi)\textbf{e}^{i+1T}_{i+1}\\
	\textbf{0}_{1\times(i+1)} & 1\\
	\end{bmatrix}\cdot\textbf{p}_L^{\phi}(\ell,\cdot),	
	\end{split}
	\label{recursive1}
	\end{align} 
	where $\textbf{p}^{\phi}_L(\ell,\cdot):=[p^{\phi}_L(\ell,q_{0}), p^{\phi}_L(\ell,q_{1}), \dots, p^{\phi}_L(\ell,q_{K})]^T$, $\mathbb{P}(\pi)$ denotes the stationary probability of a state satisfying the predicate $\pi$, $\textbf{e}^{j}_i$ denotes a row vector where the $j$th entry is one and all other entries are zeros, and
	\begin{align}
	\begin{split}
	\Pi_{i+1}=
	\begin{bmatrix}
	\mathbb{P}(\pi)  & \mathbb{P}(\lnot\pi) & 0 & \dots & 0 \\
	\mathbb{P}(\pi)  & 0 & \mathbb{P}(\lnot\pi) & \dots & 0 \\    
	\vdots  & \vdots   & \vdots & & \vdots  \\ 
	\mathbb{P}(\pi)  & 0 & 0 & \dots & \mathbb{P}(\lnot\pi) \\  
	\mathbb{P}(\pi)  & 0 & 0 & \dots & 0 \\                                                                                         
	\end{bmatrix}_{(i+1)\times(i+1)}.	
	\end{split}
	\end{align} 
	Then we have                                                           
	\begin{align}
	\begin{split}
	\textbf{p}_L^{\phi}(1,\cdot)=\begin{bmatrix}
	\Pi_{i+1} & \mathbb{P}(\lnot\pi)\textbf{e}^{i+1T}_{i+1}\\
	\textbf{0}_{1\times(i+1)} & 1\\
	\end{bmatrix}^{L-1}\cdot\textbf{p}_L^{\phi}(L,\cdot).	
	\end{split}
	\label{recursive1}
	\end{align}
	The computation burden mainly lies in computing $\Pi_{i+1}^{L-1}$. \\
	For $\phi'=\Box\Diamond_{\le i+1}\pi$, the recursion becomes the following ($K=i+2$):
	\begin{align}
	\begin{split}
	\textbf{p}_L^{\phi'}(1,\cdot)=\begin{bmatrix}
	\Pi_{i+2} & \mathbb{P}(\lnot\pi)\textbf{e}^{i+2T}_{i+2}\\
	\textbf{0}_{1\times(i+2)} & 1\\
	\end{bmatrix}^{L-1}\cdot\textbf{p}_L^{\phi'}(L,\cdot),	
	\end{split}
	\label{recursive2}
	\end{align}
	where 
	\begin{align}
	\begin{split}
	\Pi_{i+2}=             
	\begin{bmatrix}
	\Pi_{i+1} & \mathbb{P}(\lnot\pi)\textbf{e}^{i+1T}_{i+1} \\
	\mathbb{P}(\pi)\textbf{e}^{1}_{i+1}  & 0 \\                                                                                         
	\end{bmatrix}.	
	\end{split}
	\end{align} 
	
	If we reuse the results of $\Pi_{i+1}^{L-1}$ when computing $\Pi_{i+2}^{L-1}$, then the extra cost of computing (\ref{recursive2}) is only $O(LK)$.
\end{remark} 

\subsection{Computation of Information Gain for Prior Probability Distributions Governed by DTMCs}
\label{gain_B}
In this subsection, we present an algorithm to compute the information gain for prior probability distributions governed by discrete-time Markov chains, which can be seen as generalizations of stationary prior probability distributions.

\begin{definition}
	A \textit{discrete-time Markov chain} (DTMC) is defined by a tuple $\mathcal{M}= (S, S_0, P, \mathcal{AP},\mathcal{L})$, where $S=\{s^1,$ $s^2,\dots,s^H\}$ is a finite set of states, $S_0\subset S$ is the initial set of states, $P: S\times S$ $\rightarrow[0,1]$ is the transition probability, $P(s,$ $s')\ge0$ for all $s, s'\in S$ and $\sum_{s'\in S}P(s,s')=1$, $\mathcal{AP}$ is a set of atomic predicates, and $\mathcal{L}: S\rightarrow2^{\mathcal{AP}}$ is a labeling function.
\end{definition} 

In the following, the DTMC that governs the prior probability distribution will be referred to as the \textit{prior} DTMC. 

\begin{definition}[Product Automaton]
	Let $\mathcal{M}= (S, S_0, P, $ $\mathcal{AP},\mathcal{L})$ be a DTMC and $\mathcal{A}^{\phi}=(\mathcal{Q},q_0,2^{\mathcal{AP}},\delta,Acc^{\phi})$ be a DFA. The product automaton $\mathcal{M}^{\phi}_{\rm{p}}:=\mathcal{M}\otimes\mathcal{A}^{\phi}=(S_{\rm{p}},S_{0\rm{p}},P_{\rm{p}},\mathcal{L}_{\rm{p}},Acc_{\rm{p}}^{\phi})$ is a tuple such that
	\begin{itemize}
		\item $S_{\rm{p}}=S\times\mathcal{Q}$ is a finite set of states;
		\item $S_{0\rm{p}}$ is the initial set of states where for each $s_{0\rm{p}}=(s^i,q)\in S_{0\rm{p}}$, $s^i\in S_0$, $q=\delta(q_0, \mathcal{L}(s^i))$;	
		\item $P_{\rm{p}}((s,q),(s',q'))=\begin{cases}
		P(s,s')  ~~\mbox{if}~q'=\delta(q, \mathcal{L}(s'));\\ 
		0 ~~~~~~~~~~~~\mbox{otherwise};
		\end{cases}$\\
		\item $\mathcal{L}_{\rm{p}}((s,q))=\{q\}$ is a labeling function; and 
		\item $Acc^{\phi}_{\rm{p}}=S\times Acc^{\phi}$ is a finite set of accepting states.
	\end{itemize}
\end{definition} 

\begin{proposition} 
	For a (co-)safe pLTL formula $\phi$ and a \textit{prior} DTMC $\mathcal{M}$, Algorithm \ref{compute2} returns $\mathcal{I}(\mathcal{F}_L,\mathcal{\bar{F}}^{\phi}_L)$.
	\label{computeI2}
\end{proposition} 

\textbf{proof}     
See Appendix. 

\begin{remark}
	The time complexity of Algorithm \ref{compute2} is $O(LK^2H^2)$, where $H$ is the number of states of $\mathcal{M}$. By storing the results of computation for one (co-)safe pLTL formula, the extra cost of computation for another (co-) safe pLTL formula with the same predicates, same temporal operators but different temporal parameters is $O(LKH)$.
\end{remark}

\begin{algorithm}[t] 
	\caption{Information gain for prior probability distributions governed by DTMCs.}                             
	\label{compute2}
	\begin{algorithmic}[1]
		\State \textbf{Input:} $\mathcal{S}_L=\{\hat{s}^1_{1:L}, \dots, \hat{s}^m_{1:L}\}, \phi, \mathcal{M}$
		\State $\beta\gets\mathbb{\bar{P}}_{\mathcal{S}_L,\phi}$
		\State Obtain the product automaton $\mathcal{M}_{\rm{p}}^{\phi}$ (if $\phi$ is syntactically co-safe) or $\mathcal{M}_{\rm{p}}^{\lnot\phi}$ (if $\phi$ is syntactically safe) 
		\For{$k=0$ to $K$}  
		\State \textbf{if} $q_{k}\in Acc^{\phi}$ (resp. $Acc^{\lnot\phi}$) \textbf{then} 
		\State ~~~~~$\forall i\in[1, H]$, $p(L,s^i,q_{k})\gets1$ 
		\State \textbf{else} $\forall i\in[1, H]$, $p(L,s^i,q_{k})\gets0$ 
		\State \textbf{end if}                               
		\EndFor		     
		\For{all $j$, $k$, $i_1$, $i_2$}
		\State $\textbf{C}^{\rm{p}}_{j,k}(i_1, i_2)\gets P_{\rm{p}}((s^{i_1},q_j),(s^{i_2},q_k))$
		\EndFor
		\State \textbf{for} $\ell=L$ to 2, $j=0$ to $K$, $i_1=1$ to $H$ \textbf{do}                                                          
		\State $p(\ell-1,s^{i_1}, q_{j})\gets\sum_{k=0}^{K}\sum^{H}_{i_2=1}\textbf{C}^{\rm{p}}_{j,k}(i_1, i_2)\cdot p(\ell,s^{i_2}, q_{k})$  
		\State \textbf{end for} 
		\State \textbf{if} $\phi$ is syntactically co-safe \textbf{then} 
		\State ~~~~~$\gamma\gets\sum\limits_{s^i\in S_0}p_{\rm{int}}(s^i)p(1,s^i, q_{0})$	
		\State \textbf{else} $\gamma\gets\sum\limits_{s^i\in S_0}p_{\rm{int}}(s^i)(1-p(1,s^i, q_{0}))$	
		\State \textbf{end if}   		
		\State return $\mathcal{I}=\frac{1}{L}(\beta\log(\frac{\beta}{\gamma})+(1-\beta)\log(\frac{1-\beta}{1-\gamma}))$	
	\end{algorithmic}
\end{algorithm}	

\section{Information-Guided pLTL Inference}
In this section, we present the algorithm for our information-guided temporal logic inference approach. Computing the information gain for a (co-)safe pLTL formula is the essential step for the inference. For a (co-)safe pLTL formula that contains Boolean connectives (i.e., conjunctions and disjunctions), the computation of the information gain is expensive as the number of states of the DFA is exponential with respect to the size of the formula \cite{CalinBook}. 

To reduce the computational cost, we adopt a heuristic method to solve Problem \ref{problem1} for a subset of (co-)safe pLTL formulas. At first, we infer a set of (co-)safe pLTL formulas that do not contain Boolean connectives. Such pLTL formulas will be referred to as \textit{primitive} pLTL formulas. Each of the inferred primitive pLTL formula, denoted as $\phi^1_j$, represents a \textit{subpattern} that provides the maximal information gain (computed by the algorithms in Section \ref{gain}) with a certain temporal operator. We then form a \textit{pattern} consisting of the subpatterns in the form of $\phi^1=\phi_1^1\wedge\phi_2^1\wedge\dots\wedge\phi_q^1$. If $\phi^1$ does not satisfy the coverage constraint (i.e., $\mathbb{\bar{P}}_{\mathcal{S}_L, \phi^1}<p_{\rm{th}}$), we remove the trajectories in $\mathcal{S}_L$ that satisfy $\phi^1$ and find another pattern $\phi^2$ from the remaining trajectories. And the same procedure continues until all the obtained patterns already cover $p_{\rm{th}}$ portion of the trajectories in $\mathcal{S}_L$. The formulas $\phi^j$ (representing different patterns) are then connected in disjunction to form the final inferred formula. 

The proposed information-guided temporal logic inference approach is outlined in Algorithm \ref{algMain}. Initially, $\mathcal{S}_L$ is given as the dataset $\{\hat{s}^1_{1:L}, \dots, \hat{s}^m_{1:L}\}$, $\mathcal{\hat{B}}_L$ is given as $\emptyset$, $\phi$ is given as $\top$, and $\beta$ is given as 0. We define a \textit{primitive template} as a primitive pLTL formula with fixed temporal operator, fixed predicate structure and undetermined parameters (including the temporal parameters and the parameters in the predicates). We select a set $\mathcal{P}$ of primitive templates, and for each primitive template $\phi_k\in\mathcal{P}$ and the parameter vector $\theta_k$, we solve the following optimization problem:
\begin{align}
\begin{split}
&\max\limits_{\theta_k\in\Theta_k}\mathcal{I}(\mathcal{F}_L,\mathcal{\bar{F}}^{\phi_k(\theta_k)}_L)
\textrm{subject~to~} \mathbb{\bar{P}}_{\mathcal{S}_L, \phi_k(\theta_k)}\ge\hat{p}_{\rm{th}},
\end{split}
\label{oneStep}
\end{align}
where $\Theta_k$ is the set of parameter vectors for $\phi_k$, $\hat{p}_{\rm{th}}\in(0, p_{\rm{th}}]$ is a hyperparameter representing a coverage threshold for each pattern. $\hat{p}_{\rm{th}}$ should not be set too large (e.g. $\hat{p}_{\rm{th}}=p_{\rm{th}}$) as each pattern may not cover a large portion of the trajectories in $\mathcal{S}_L$. $\hat{p}_{\rm{th}}$ should not be set too small as well, as it may lead to a very long formula consisting of many patterns in the end and violating the size constraint.

To solve the constrained non-convex optimization problem (\ref{oneStep}), we use particle swarm optimization (PSO) \cite{FJZ2013} to optimize the parameter vector $\theta_k$ for each primitive template $\phi_k$ for the following unconstrained non-convex optimization problem transformed from (\ref{oneStep}):\\
\begin{align}
\begin{split}
&\min\limits_{\theta_k\in\Theta_k}-\mathcal{I}(\mathcal{F}_L,\mathcal{\bar{F}}^{\phi_k(\theta_k)}_L)+G(\mathbb{\bar{P}}_{\mathcal{S}_L, \phi_k(\theta_k)}), 
\end{split}
\label{oneStep2}
\end{align}
where 
\[                                   
\begin{split}
G(x)=                                                            
\begin{cases}
\varrho\cdot(\hat{p}_{\rm{th}}-x)  ~~~~~~~~\mbox{if}~x\le\hat{p}_{\rm{th}};\\ 
0 ~~~~~~~~~~~~~~~~~~~~~\mbox{otherwise},
\end{cases}
\end{split}
\]
and $\varrho$ is a large positive number. As $\phi_k(\theta_k)$ does not contain Boolean connectives, $\mathcal{I}(\mathcal{F}_L,\mathcal{\bar{F}}^{\phi_k(\theta_k)}_L)$ can be efficiently computed by Algorithm \ref{compute} or Algorithm \ref{compute2}.

After we obtain the formula $\phi_k^{\ast}$ by solving (\ref{oneStep2}) for each primitive template, we rank $\phi_k^{\ast}$ according to $\mathcal{I}(\mathcal{F}_L,\mathcal{\bar{F}}^{\phi_k^{\ast}}_L)$ (from the highest to the lowest) and obtain $\{\hat{\phi}_j^{\ast}\}^{\vert\mathcal{P}\vert}_{j=1}$. For the formula $\hat{\phi}_1^{\ast}$, we further increase the information gain by adding conjunctions between $\hat{\phi}_1^{\ast}$ and the next formula $\hat{\phi}_j^{\ast}$ (with $j$ selected in increasing order from $2,3,\dots,\vert\mathcal{P}\vert$) that satisfies two conditions: (1) $\mathcal{I}(\mathcal{F}_L,\mathcal{\bar{F}}^{\hat{\phi}_1^{\ast}\wedge\hat{\phi}_j^{\ast}}_L)\ge\alpha\mathcal{I}(\mathcal{F}_L,\mathcal{\bar{F}}^{\hat{\phi}_1^{\ast}}_L)$; (2) $\mathbb{\bar{P}}_{\mathcal{S}_L,\hat{\phi}_1^{\ast}\wedge\hat{\phi}_j^{\ast}}\ge\hat{p}_{\rm{th}}$. Condition (1) guarantees that information gain is increased by adding each conjunction, with larger $\alpha$ $(\alpha>1)$ representing higher information gain required for adding each conjunction, condition (2) guarantees that the coverage threshold we set is still valid for this pattern by adding each conjunction. 

To reduce the computational cost in computing the information gain $\mathcal{I}(\mathcal{F}_L,\mathcal{\bar{F}}^{\phi}_L)$ when $\phi$ contains Boolean connectives, we simulate a set of trajectories $\mathcal{\hat{B}}_L$ based on the prior probability distribution $\mathcal{F}_L$ and calculate an estimated information gain $\mathcal{\hat{I}}(\mathcal{F}_L,\mathcal{\bar{F}}^{\phi}_L)$ using the estimated probability $\mathbb{P}_{\mathcal{\hat{B}}_L,\phi}$ for checking condition (1). If a formula $\hat{\phi}_j^{\ast}$ satisfies the two conditions, then we add conjunctions between $\hat{\phi}_1^{\ast}\wedge\hat{\phi}_j^{\ast}$ and the next formula $\hat{\phi}_k^{\ast}$ that satisfies $\mathcal{\hat{I}}(\mathcal{F}_L,\mathcal{\bar{F}}^{\hat{\phi}_1^{\ast}\wedge\hat{\phi}_j^{\ast}\wedge\hat{\phi}_k^{\ast}}_L)\ge\alpha\mathcal{\hat{I}}(\mathcal{F}_L,\mathcal{\bar{F}}^{\hat{\phi}_1^{\ast}\wedge\hat{\phi}_j^{\ast}}_L)$ and $\mathbb{\bar{P}}_{\mathcal{S}_L,\hat{\phi}_1^{\ast}\wedge\hat{\phi}_j^{\ast}\wedge\hat{\phi}_k^{\ast}}\ge\hat{p}_{\rm{th}}$, and the same procedure continues until the last formula $\hat{\phi}_{\vert\mathcal{P}\vert}^{\ast}$ is checked.             

After we obtain the formula $\hat{\phi}=\hat{\phi}_1^{\ast}\wedge\hat{\phi}_j^{\ast}\wedge\dots\wedge\hat{\phi}_q^{\ast}$, if the coverage constraint is already satisfied, then the algorithm terminates. Otherwise, we remove the trajectories that satisfy $\hat{\phi}$ and repeat the process for the remaining trajectories in $\mathcal{S}^{\lnot\hat{\phi}}_L$, where $\mathcal{S}^{\lnot\hat{\phi}}_L$ denotes the set of paths in $\mathcal{S}_L$ that do not satisfy $\hat{\phi}$. The obtained formula from $\mathcal{S}^{\lnot\hat{\phi}}_L$ is connected in disjunction with $\hat{\phi}$ and the same procedure goes on until the coverage constraint is satisfied.

The time complexity of Algorithm \ref{algMain} is $O(M\vert\mathcal{P}\vert g(m))$, where $g(m)$ is the time complexity of the local PSO algorithm for each primitive template, $m$ is the number of trajectories in the dataset, $M$ is the number of iterations needed for Algorithm \ref{algMain} to terminate ($M\le\log_{1-\hat{p}_{\rm{th}}}(1-p_{\rm{th}})+1$). It can be seen that Algorithm \ref{algMain} has polynomial time complexity with respect to the size of the formula. If we set $(\vert\mathcal{P}\vert-1)\log_{1-\hat{p}_{\rm{th}}}(1-p_{\rm{th}})\le\ell_{\rm{th}}$, then the size constraint in Problem \ref{problem1} can be guaranteed. 

\begin{algorithm}[t]
	\caption{Information-Guided Temporal Logic Inference.}                            
	\label{algMain}
	\begin{algorithmic}[1]
		\State \textbf{procedure} $TLprocedure(\mathcal{S}_L, \mathcal{\hat{B}}_L, \phi, \beta)$ 
		\For{$k=1:\vert\mathcal{P}\vert$ }   	                                                              
		\State $\displaystyle\phi_k^{\ast}\gets\arg\max_{\theta_k\in\Theta_k}\mathcal{I}(\mathcal{F}_L,\mathcal{\bar{F}}^{\phi_k(\theta_k)}_L)$
		\EndFor
		\State Obtain $\{\hat{\phi}_j^{\ast}\}^{\vert\mathcal{P}\vert}_{j=1}$ by ranking $\phi_k^{\ast}$ according to $\mathcal{I}(\mathcal{F}_L,\mathcal{\bar{F}}^{\phi_k^{\ast}}_L)$ 
		\State $\hat{\phi}\gets\hat{\phi}_1^{\ast}$
		\If{$\mathcal{\hat{B}}_L=\emptyset$}
		\State Generate/simulate a set of trajectories $\mathcal{\hat{B}}_L$ based on the prior probability distribution $\mathcal{F}_L$
		\EndIf
		\For{$j=2:\vert\mathcal{P}\vert$}   
    	\State \textbf{if} $\mathcal{\hat{I}}(\mathcal{F}_L,\mathcal{\bar{F}}^{\hat{\phi}\wedge\hat{\phi}_j^{\ast}}_L)\ge\alpha\mathcal{\hat{I}}(\mathcal{F}_L,\mathcal{\bar{F}}^{\hat{\phi}}_L)$ and $\vert\mathbb{\bar{P}}_{\mathcal{S}_L,\hat{\phi}\wedge\hat{\phi}_j^{\ast}}-\mathbb{\bar{P}}_{\mathcal{S}_L,\hat{\phi}}\vert\le\epsilon$ \textbf{then} $\hat{\phi}\gets\hat{\phi}\wedge\hat{\phi}_j^{\ast}$  
		\State \textbf{end if} 
		\EndFor
		\State $\beta\gets\beta+(1-\beta)\mathbb{\bar{P}}_{\mathcal{S}_L,\hat{\phi}}$, $\phi\gets\phi\vee\hat{\phi}$ 
		\State \textbf{if} $\beta\ge p_{\rm{th}}$ \textbf{then} return $\phi$   
        \State \textbf{else} $\mathcal{S}^{\lnot\hat{\phi}}_L\gets\{\hat{s}^i_{1:L}~\vert~\hat{s}^i_{1:L}\in\mathcal{S}_L, \hat{s}^i_{1:L}\not\models\hat{\phi}\}$	        	                             
		\State ~~~~~~$\phi\gets TLprocedure(\mathcal{S}^{\lnot\hat{\phi}}_L, \mathcal{\hat{B}}_L, \phi, \beta)$
		\State \textbf{end if}  
		\State Return $\phi$ 		
		\State\textbf{end procedure}
	\end{algorithmic}
\end{algorithm}	                                                         

\section{Implementation}

\subsection{Case I: Explaining Anomalous Patterns}
\begin{figure}[th]
	\centering
	\includegraphics[width=10cm]{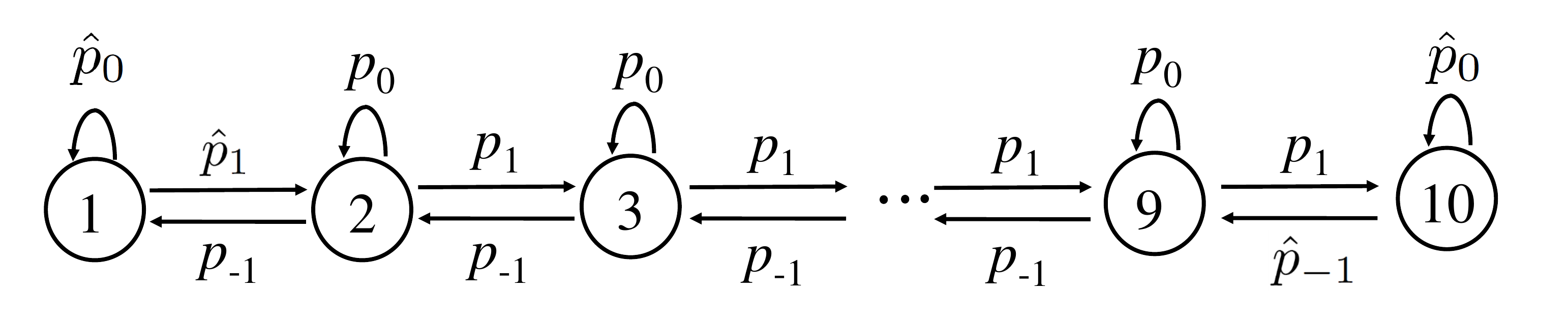}\caption{The \textit{prior} DTMC $\mathcal{M}$ in Case I and II.}
	\label{DTMC}
\end{figure}
In the first case study,  we implement the inference approach in identifying and explaining the anomalous patterns ``hidden'' in a set of trajectories. The \textit{prior} DTMC is as shown in Fig. \ref{DTMC}, and it is parametrized with the following parameters: $p_0=p_{-1}=p_{1}=\frac{1}{3}, \hat{p}_0=\hat{p}_{1}=\hat{p}_{-1}=0.5$. The initial probability distribution is set as $p_{\rm{int}}(s^i)=0.1$ where $s^i=i$, $(i=1,2,\dots,10)$. We inject two anomalous patterns by adding the following constraints when simulating 100 trajectories from the \textit{prior} DTMC: for each trajectory of the first 60 trajectories, the minimal number for the first 5 time indices are constrained to the number 1 or 2, and the two numbers from time index 51 to 52 are constrained to the number 9 or 10, for each trajectory of the last 60 trajectories, the minimal number during the 10 time indices from 71 to 80 are constrained to the number 1 or 2 (for 20 trajectories the two anomalous patterns both exist). We set $p_{\rm{th}}=0.9, \hat{p}_{\rm{th}}=0.5, \ell_{\rm{th}}=10, \alpha=1.5, \varrho=1000$. We infer a pLTL formula from a set of primitive templates \cite{Bombara2016}
\begin{align}
\mathcal{P}=\{\Box_{I}\pi, \Diamond_{I}\pi, \Box\Diamond_{I}\pi, \Diamond\Box_{I}\pi\},
\label{template}
\end{align}
where $I$ can be $\ge i$, $\le i$ or $(\ge i_1, \le i_2)$ (here $i_1<i_2$, $i, i_1, i_2\in\mathbb{T}$), $\pi$ is an atomic predicate in the form of $x\ge a$ or $x\le a$, where $a\in[1,10]$ is an integer parameter. In the first iteration, the primitive pLTL formula with the highest information gain is $\phi^{1\ast}_1=\Box_{\ge51,\le52}(x\ge9)$ (see Tab. \ref{results_I_1}). To add conjunctions with other inferred primitive pLTL formulas, we simulate a set of 100 trajectories from the \textit{prior} DTMC and find only $\phi^{1\ast}_4=\Diamond_{\le5}(x\le2)$ satisfying the conditions in line 11 of Algorithm \ref{algMain} to be added in conjunction with $\phi^{1\ast}_1$. In comparison, $\phi^{1\ast}_3$ also has high information gain as a primitive pLTL formula, but the estimated information gain $\mathcal{\hat{I}}(\mathcal{F}_L,\mathcal{\bar{F}}^{\phi_1^{1\ast}\wedge\phi_3^{1\ast}}_L)$ is computed the same as $\mathcal{\hat{I}}(\mathcal{F}_L,\mathcal{\bar{F}}^{\phi_1^{1\ast}}_L)$, as $\phi_3^{1\ast}$ essentially describes the same subpattern as $\phi_1^{1\ast}$. As $\mathbb{\bar{P}}_{\mathcal{S}_L,\phi^{1\ast}_1\wedge\phi^{1\ast}_4}=0.61 <p_{\rm{th}}$, the algorithm continues for the second iteration and finds the second pattern $\phi^{2\ast}_4=\Diamond_{\ge71,\le80}(x\le2)$ from the remaining 39 trajectories that do not satisfy $\phi^{1\ast}_1\wedge\phi^{1\ast}_4$. Then the algorithm terminates with $\mathbb{\bar{P}}_{\mathcal{S}_L,(\phi^{1\ast}_1\wedge\phi^{1\ast}_4)\vee\phi^{2\ast}_4}=0.61+0.39\cdot1>p_{\rm{th}}$. 
Therefore, the final inferred pLTL formula is
\[
\begin{split}
\phi^{\ast}=(\Box_{\ge51,\le52}(x\ge9)\wedge\Diamond_{\le5}(x\le2))\vee(\Diamond_{\ge71,\le80}(x\le2)).
\end{split}
\]

\begin{table}
	\centering \caption{Results in First Iteration of Case Study I}         
	\begin{tabular}{|c|c|c|c|c|c|c|c|}                                             
		\hline 
		pLTL formula & $\mathbb{\bar{P}}_{\mathcal{S}_L,\phi^{\ast}}$ & $\mathbb{P}_{\mathcal{B}_L,\phi^{\ast}}$ & $\mathcal{I}(\mathcal{F}_L,\mathcal{\bar{F}}^{\phi^{\ast}}_L)$   
		\\ \hline \tabincell{c}{$\phi^{1\ast}_1=\Box_{\ge51,\le52}(x\ge9)$} &  0.63 &  0.1429   &  0.0062
		\\ \hline \tabincell{c}{$\phi^{1\ast}_2=\Box_{\ge3,\le4}(x\le3)$} &  0.69 &  0.2609   & 0.004
		\\ \hline \tabincell{c}{$\phi^{1\ast}_3=\Diamond_{\ge50,\le51}(x\ge9)$} & 0.67 &  0.2143   &  0.0048
		\\ \hline \tabincell{c}{$\phi^{1\ast}_4=\Diamond_{\le5}(x\le2)$} & 0.76  &  0.2877  &  0.0048
		\\ \hline \tabincell{c}{$\phi^{1\ast}_5=\Box\Diamond_{\le51}(x\ge10)$} & 0.59  &  0.2906    &  0.0019
		\\ \hline \tabincell{c}{$\phi^{1\ast}_6=\Box\Diamond_{\le96}(x\le2)$}  & 0.99  &  0.7781   &  0.0021
		\\ \hline \tabincell{c}{$\phi^{1\ast}_7=\Diamond\Box_{\le2}(x\ge9)$}  & 0.85 &  0.7167   &  0.000496
		\\ \hline \tabincell{c}{$\phi^{1\ast}_8=\Diamond\Box_{\le3}(x\le3)$}  & 0.99  &  0.7808   &  0.002	
		\\ \hline
	\end{tabular}   
	\label{results_I_1}                        
\end{table}

\subsection{Case II: Explaining Pattern Changes}
In the second case study, we implement the inference approach in inferring a pLTL formula to explain the pattern changes from the \textit{prior} DTMC to the DTMC \textit{at present}. We simulate 100 trajectories from the DTMC \textit{at present} as shown in Fig. \ref{DTMC} with the following parameters: $p_0=0.2, p_{-1}=0.6, p_{1}=0.2, \hat{p}_0=\hat{p}_{1}=\hat{p}_{-1}=0.5$. The \textit{prior} DTMC has the same structure as that of the DTMC \textit{at present}, but with different parameters. In Scenario (a), we set $p^{\rm{a}}_0=p^{\rm{a}}_{-1}=p^{\rm{a}}_{1}=\frac{1}{3}, \hat{p}^{\rm{a}}_0=\hat{p}^{\rm{a}}_{1}=\hat{p}^{\rm{a}}_{-1}=0.5$; in Scenario (b), we set $p^{\rm{b}}_0=0.05, p^{\rm{b}}_{-1}=0.9, p^{\rm{b}}_{1}=0.05, \hat{p}^{\rm{b}}_0=\hat{p}^{\rm{b}}_{1}=\hat{p}^{\rm{b}}_{-1}=0.5$. In both scenarios, we use the same hyperparameters and the same set of primitive templates as in Case Study I.

In Scenario (a), the final inferred pLTL formula is
\[
\begin{split}
\phi^{\ast}_{\textrm{a}}=&(\Box\Diamond_{\le22}(x\le1)\wedge\Diamond\Box_{\le78}(x\le5))\vee\\
&(\Box_{\ge53,\le99}(x\le4)\wedge\Box\Diamond_{\le32}(x\le1)).
\end{split}
\]

In Scenario (b), the final inferred pLTL formula is
\[
\begin{split}
\phi^{\ast}_{\textrm{b}}=&(\Diamond_{\ge16,\le99}(x\ge5))\vee(\Diamond_{\ge13,\le99}(x\ge4))
\vee(\Diamond_{\ge14,\le30}(x\ge3)).
\end{split}
\]
It can be seen that the predicates in $\phi^{\ast}_{\textrm{a}}$ are all in the form of $x\le a$, as the DTMC \textit{at present} has the trend towards smaller numbers in comparison with the \textit{prior} DTMC in Scenario (a), while the predicates in $\phi^{\ast}_{\textrm{b}}$ are all in the form of $x\ge a$, as the \textit{prior} DTMC in Scenario (b) has the trend towards even smaller numbers in comparison with the DTMC \textit{at present}.

\subsection{Case III: Explaining Policies of MDPs}

\begin{figure}[th]
	\centering
	\includegraphics[width=12cm]{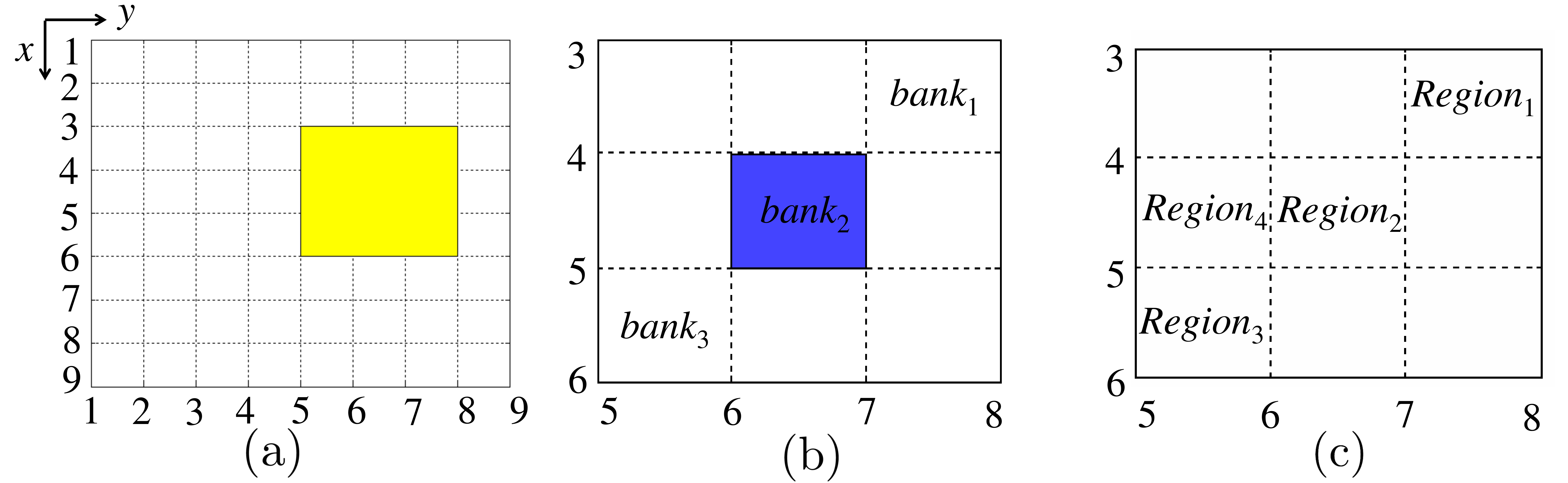}\caption{The game setup and inferred regions in Case III.}                    
	\label{regionIII}
\end{figure}

\begin{table*}[t]
	\centering \caption{Results in First Iteration of Second Task of Case Study III}          
	\begin{tabular}{|c|c|c|c|c|c|c|c|}
		\hline 
		inferred pLTL formula & $\mathbb{\bar{P}}_{\mathcal{S}_L,\psi^{\ast}}$ & $\mathbb{P}_{\mathcal{B}_L,\psi^{\ast}}$ & $\mathcal{I}(\mathcal{F}_L,\mathcal{\bar{F}}^{\psi^{\ast}}_L)$  & $\eta(\psi^{\ast})$
		\\ \hline \tabincell{c}{$\psi^{\ast}_{b1,1}=\Box\big(((4\le x\le6)\wedge(7\le y\le8)$\\ $\wedge(b_{\textrm{cop}}=bank_1))\rightarrow\Box_{\le5}((4\le x\le5)\wedge(6\le y\le7))\big)$} &  0.9844 & $1.69\cdot 10^{-5}$ & 0.1073 & 0.0007
		\\ \hline \tabincell{c}{$\psi^{\ast}_{b1,2}=\Box\big(((3\le x\le6)\wedge(5\le y\le8)$\\ $\wedge(b_{\textrm{cop}}=bank_1))\rightarrow\Diamond_{\le2}((4\le x\le5)\wedge(6\le y\le7))\big)$}  &  0.8438 & $5.58\cdot 10^{-16}$ & 0.292 & 0.9771
		\\ \hline \tabincell{c}{$\psi^{\ast}_{b2,1}=\Box\big(((3\le x\le4)\wedge(5\le y\le8)$\\ $\wedge(b_{\textrm{cop}}=bank_2))\rightarrow\Box_{\le5}((3\le x\le4)\wedge(7\le y\le8))\big)$} &  0.9323 & $5.58\cdot 10^{-16}$& 0.325 & 0.0105
		\\ \hline \tabincell{c}{$\psi^{\ast}_{b2,2}=\Box\big(((3\le x\le6)\wedge(5\le y\le8)$\\ $\wedge(b_{\textrm{cop}}=bank_2))\rightarrow\Diamond_{\le2}((4\le x\le6)\wedge(5\le y\le6))\big)$} &  0.9323   &  $7.8\cdot 10^{-16}$ & 0.3218  & 0.9784
		\\ \hline \tabincell{c}{$\psi^{\ast}_{b3,1}=\Box\big(((4\le x\le6)\wedge(7\le y\le8)$\\ $\wedge(b_{\textrm{cop}}=bank_3))\rightarrow\Box_{\le4}((4\le x\le5)\wedge(6\le y\le7))\big)$} &  0.9948       & $1.69\cdot 10^{-5}$ & 0.109 & 0.0029
		\\ \hline \tabincell{c}{$\psi^{\ast}_{b3,2}=\Box\big(((3\le x\le6)\wedge(5\le y\le8)$\\ $\wedge(b_{\textrm{cop}}=bank_3))\rightarrow\Diamond_{\le2}((4\le x\le5)\wedge(6\le y\le7))\big)$}&  0.9948 & $5.58\cdot 10^{-16}$ & 0.3491 & 0.86  
		\\ \hline
	\end{tabular}                  
	\label{caseIII}                                                    
\end{table*} 
In the third case study, we implement the inference approach on a ``cops and robbers'' game with the inferred pLTL formulas aiming to explain the policies of the robber as an agent whose behaviors are governed by an MDP. As shown in Fig. \ref{regionIII}, there are three banks locating in three $1\times1$ blocks in the state space, where the robber's goal is to reach $bank_2$ while deceiving the cop (as an adversary) to believe that the robber is trying to reach the other two banks (see \cite{deception} for details of the game). We simulate 192 trajectories with length 100, with the policies obtained from \cite{deception}, starting from the 64 different initial states with 3 different initial beliefs of the cop. We set the prior probability distribution as a stationary uniform distribution in the state space. 

In the first task, we infer a pLTL formula from the set of primitive templates (\ref{template}) by replacing $\pi$ with $\iota$, where $\iota$ is any rectangular predicate in the state space. We use the same hyperparameters as in Case Study I and II except that $p_{\rm{th}}$ is set as 0.95 as we expect the explanations of the policies to cover a higher percentage of the trajectories. The final inferred formula for the first task is
\[
\begin{split}
\phi^\ast_{\textrm{f}}=\Diamond\Box_{\le30}((3\le x\le6)\wedge(5\le y\le8)),
\end{split}
\] 
which means ``the robber will eventually reach the yellow region in Fig. \ref{regionIII} and stay there for at least 30 time indices''. 

In the second task, to further explain the policies of the robber as a response to both the current position of the robber and the cop's beliefs, we infer a pLTL formula in the causal form $\psi=\Box(\psi_{\textrm{c}} \Rightarrow\psi_{\textrm{e}})$. The cause formula is in the form of $\psi_{c}=\iota_{\textrm{y}}\wedge \pi_{\textrm{b}}$, where $\iota_{\textrm{y}}$ is a rectangular predicate of the robber's position in the yellow region (we shrink the state space to the yellow
region as $\phi^\ast_{\textrm{f}}$ indicates) and $\pi_{\textrm{b}}$ is a predicate of the belief of the cop in the form of $b_{\textrm{cop}}=bank_i$ ($i=1,2,3$). The effect formula $\psi_{\textrm{e}}$ is in the form of $\psi_{\textrm{e}}=\Box_{I}\iota_{\textrm{y}}$ or $\psi_{\textrm{e}}=\Diamond_{I}\iota_{\textrm{y}}$, where $I$ is defined in (\ref{template}). 

There is a caveat though for pLTL formulas of the causal form $\psi=\Box(\psi_{\textrm{c}}\Rightarrow\psi_{\textrm{e}})$: $\psi$ is satisfied by a trajectory if $\psi_{\textrm{c}}$ is never satisfied by the trajectory at any time index. To counteract this, we define the truth factor $\eta(\psi)$ as the proportion of the states in all the trajectories that satisfy the cause formula $\psi_{\textrm{c}}$, and add $\eta(\psi)>0$ as an extra constraint in the optimization. Besides, we modify Algorithm \ref{algMain} for pLTL formulas of the causal form as follows: (i) in line 5 of Algorithm \ref{algMain}, we rank the obtained primitive formulas based on $\mathcal{I}(\mathcal{F}_L,\mathcal{\bar{F}}^{\psi_k^{\ast}}_L)+\eta(\psi_k^{\ast})$ instead of  $\mathcal{I}(\mathcal{F}_L,\mathcal{\bar{F}}^{\psi_k^{\ast}}_L)$; (ii) in line 16-17 of Algorithm \ref{algMain}, if the inferred formula $\hat{\psi}=\Box(\hat{\psi}_{\textrm{c}}\Rightarrow\hat{\psi}_{\textrm{e}})$ does not satisfy the coverage constraint, then we infer another primitive formula $\hat{\psi}^2_{\textrm{e}}$ that is connected in disjunction with $\hat{\psi}_{\textrm{e}}$, and the same procedure continues until the coverage constraint is satisfied. We use the same hyperparameters as in the first task and infer three pLTL formulas corresponding to the three different beliefs of the cop. After the first iteration, the inferred pLTL formulas for the three different beliefs of the cop are $\psi^{\ast}_{b1,2}$, $\psi^{\ast}_{b2,2}$ and $\psi^{\ast}_{b3,2}$, as shown in Tab. \ref{caseIII}. 
As $\psi^{\ast}_{b3,2}$ already satisfies $\mathbb{\bar{P}}_{\mathcal{S}_L,\psi^{\ast}_{b3,2}}\ge p_{\rm{th}}$, so the final inferred formula for the belief $b_{\textrm{cop}}=bank_3$ is $\psi^{\ast}_{b3}=\psi^{\ast}_{b3,2}$. For the other two beliefs, after another iteration, the final inferred formulas are as follows:\\ 
$\psi^{\ast}_{b1}=\Box\Big(((3\le x\le6)\wedge(5\le y\le8)\wedge(b_{\textrm{cop}}=bank_1))\rightarrow\big(\Diamond_{\le2}((4\le x\le5)\wedge(6\le y\le7))\vee\Diamond_{\le2}((5\le x\le6)\wedge(5\le y\le6))\big)\Big)$,\\
$\psi^{\ast}_{b2}=\Box\Big(((3\le x\le6)\wedge(5\le y\le8)\wedge(b_{\textrm{cop}}=bank_2))\rightarrow\big(\Diamond_{\le2}((4\le x\le6)\wedge(5\le y\le6))\vee\Diamond_{\le2}((3\le x\le4)\wedge(7\le y\le8))\big)\Big)$.

$\psi^{\ast}_{b1}$, $\psi^{\ast}_{b2}$ and $\psi^{\ast}_{b3}$ mean the following: if the robber is in the yellow region and the cop believes that the robber tries to reach $bank_1$, then the robber will eventually go to $Region_2$ or $Region_3$ in 2 time steps; if the robber is in the yellow region and the cop believes that the robber tries to reach $bank_2$, then the robber will go to $Region_3$ or $Region_4$ or $Region_1$ in 2 time steps; if the robber is in the yellow region and the cop believes that the robber tries to reach $bank_3$, then the robber will eventually go to $Region_2$ in 2 time steps. 

\section{Conclusion}
We have proposed an approach to extract interpretable and informative knowledge in the form of pLTL formulas from data.
For future work, the computational methods of the information gain can be developed for more general forms of temporal logic formulas, and with other types of prior probability distributions. We will also consider other methods to solve the inference problem that is either more computationally efficient or requiring fewer data.

\section*{APPENDIX}

\textbf{Proof of Proposition \ref{exp}}:\\
From Definition \ref{know} and Definition \ref{KL}, we have
\[                           
\begin{split}
\mathcal{I}(\mathcal{F}_L,\mathcal{\bar{F}}^{\phi}_L)=&\frac{D_{\rm{KL}}(\mathcal{\bar{F}}^{\phi}_L\vert\vert\mathcal{F}_L)}{L}\\
= &
\frac{1}{L}\sum\limits_{s_{1:L}\in\mathcal{B}_L, s_{1:L}\models\phi}\mathcal{\bar{F}}^{\phi}_L(s_{1:L})\log\frac{\mathcal{\bar{F}}^{\phi}_L(s_{1:L})}{\mathcal{F}_L(s_{1:L})}\\&+
\frac{1}{L}\sum\limits_{s_{1:L}\in\mathcal{B}_L, s_{1:L}\not\models\phi}\mathcal{\bar{F}}^{\phi}_L(s_{1:L})\log\frac{\mathcal{\bar{F}}^{\phi}_L(s_{1:L})}{\mathcal{F}_L(s_{1:L})}\\
= &
\frac{1}{L}\sum\limits_{s_{1:L}\in\mathcal{B}_L, s_{1:L}\models\phi}\frac{\mathcal{F}_L(s_{1:L})\mathbb{\bar{P}}_{\mathcal{S}_L,\phi}}{\mathbb{P}_{\mathcal{B}_L,\phi}}\log\frac{\mathbb{\bar{P}}_{\mathcal{S}_L,\phi}}{\mathbb{P}_{\mathcal{B}_L,\phi}}\\&+
\frac{1}{L}\sum\limits_{s_{1:L}\in\mathcal{B}_L, s_{1:L}\not\models\phi}\frac{\mathcal{F}_L(s_{1:L})\mathbb{\bar{P}}_{\mathcal{S}_L,\lnot\phi}}{\mathbb{P}_{\mathcal{B}_L,\lnot\phi}}\log\frac{\mathbb{\bar{P}}_{\mathcal{S}_L,\lnot\phi}}{\mathbb{P}_{\mathcal{B}_L,\lnot\phi}}.
\end{split}
\]
As $\sum\limits_{s_{1:L}\in\mathcal{B}_L, s_{1:L}\models\phi}\mathcal{F}_L(s_{1:L})=\mathbb{P}_{\mathcal{B}_L,\phi}$,
we have
\[                           
\begin{split}
\mathcal{I}(\mathcal{F}_L,\mathcal{\bar{F}}^{\phi}_L) 
= &
\frac{\mathbb{\bar{P}}_{\mathcal{S}_L,\phi}}{L}\log\frac{\mathbb{\bar{P}}_{\mathcal{S}_L,\phi}}{\mathbb{P}_{\mathcal{B}_L,\phi}}+
\frac{\mathbb{\bar{P}}_{\mathcal{S}_L,\lnot\phi}}{L}\log\frac{\mathbb{\bar{P}}_{\mathcal{S}_L,\lnot\phi}}{\mathbb{P}_{\mathcal{B}_L,\lnot\phi}}\\
=&\frac{1}{L}(\mathbb{\bar{P}}_{\mathcal{S}_L,\phi}\log(\mathbb{\bar{P}}_{\mathcal{S}_L,\phi})-\mathbb{\bar{P}}_{\mathcal{S}_L,\phi}\log(\mathbb{P}_{\mathcal{B}_L,\phi})\\&+\mathbb{\bar{P}}_{\mathcal{S}_L,\lnot\phi}\log(\mathbb{\bar{P}}_{\mathcal{S}_L,\lnot\phi})-\mathbb{\bar{P}}_{\mathcal{S}_L,\lnot\phi}\log(\mathbb{P}_{\mathcal{B}_L,\lnot\phi})).
\end{split}
\]

\textbf{Proof of Proposition \ref{computeI1}}:\\
For a syntactically co-safe pLTL formula $\phi$, we use $p_L^{\phi}(\ell,q_{k})$ $(\ell\le L)$ to denote the probability of extending a trajectory of length $\ell$ to length $L$, with the state of the DFA $\mathcal{A}^{\phi}$ at time index $\ell$ being the state $q_{k}$, such that the extended trajectory of length $L$ strongly satisfies $\phi$.
Then we have
\[
\begin{split}
p_L^{\phi}(L,q_{k})=                                                            
\begin{cases}
1  ~~~~~~~~~~~~~~~\mbox{if}~q_k\in Acc^{\phi};\\ 
0 ~~~~~~~~~~~~~~~~\mbox{otherwise}.
\end{cases}
\end{split}
\]
We can compute $p_L^{\phi}(\ell,q_{k})$ recursively as follows:
\begin{align}
\begin{split}
\begin{bmatrix}
p_L^{\phi}(\ell-1,q_{0})\\
\vdots\\
p_L^{\phi}(\ell-1,q_{K})\\	
\end{bmatrix}=
\begin{bmatrix}
c_{0,0} & c_{0,1} & \dots & c_{0,K}\\                                                                                                
\vdots  & \vdots & \vdots & \vdots \\
c_{K,0} & c_{K,1} & \dots & c_{K,K}\\
\end{bmatrix}
\begin{bmatrix}
p_L^{\phi}(\ell,q_{0})\\
\vdots\\
p_L^{\phi}(\ell,q_{K})\\
\end{bmatrix},	
\end{split}
\label{recursive}
\end{align}
where $c_{j,k}$ is the probability of transitioning from $q_{j}$ to $q_{k}$ (which can be calculated from $\mathcal{F}_L$).
Finally, we have $\mathbb{P}_{\mathcal{B}_L,\phi}=p_L^{\phi}(0,q_{0})$.

	For a syntactically safe pLTL formula $\phi$, we use $p_L^{\lnot\phi}(\ell;q_{k})$ $(\ell\le L)$ to denote the probability of extending a path of time length $\ell$ to time length $L$, with the state of the DFA $\mathcal{A}^{\lnot\phi}$ at time instant $\ell$ being the state $q_{k}$, such that the extended path of time length $L$ strongly violates $\phi$ (i.e. strongly satisfies $\lnot\phi$). Then we have the following:
\[
\begin{split}
p_L^{\lnot\phi}(L;q_{k})=                                                            
\begin{cases}
1  ~~~~~~~~~~~~~~~\mbox{if}~q_k\in Acc^{\lnot\phi};\\ 
0 ~~~~~~~~~~~~~~~~\mbox{otherwise};
\end{cases}
\end{split}
\]
We can compute $p_L^{\lnot\phi}(\ell;q_{k})$ recursively using (\ref{recursive}) by replacing each $\phi$ with $\lnot\phi$. Finally, we have
$\mathbb{P}_{\mathcal{B}_L,\phi}=1-p_L^{\lnot\phi}(0;q_{0})$. 

With $\mathbb{P}_{\mathcal{B}_L,\phi}$, we can calculate $\mathcal{I}(\mathcal{F}_L,\mathcal{\bar{F}}^{\phi}_L)$ according to Proposition \ref{exp}.  

\textbf{Proof of Proposition \ref{computeI2}}:\\
For a syntactically co-safe pLTL formula $\phi$, we use $p^{\phi}_L(\ell,s^i,q_{k})$ to denote the probability of extending a trajectory of length $\ell$ to length $L$, with the state of the product automaton $\mathcal{M}^{\phi}_{\rm{p}}$ (of $\mathcal{M}$ and $\mathcal{A}^{\phi}$) at time index $\ell$ being the state $(s^i, q_{k})$, such that the extended trajectory of length $L$ strongly satisfies $\phi$.                         
	Then we have the following:
	\[
	\begin{split}
	p^{\phi}_L(L,s^i,q_{k})=                                                            
	\begin{cases}
	1  ~~~~~~~~~~~~~~~\mbox{if}~q_k\in Acc^{\phi};\\ 
	0 ~~~~~~~~~~~~~~~~\mbox{otherwise}.
	\end{cases}
	\end{split}
	\]
	We can compute $p^{\phi}_L(\ell,s^i,q_{k})$ recursively as follows:	
	\begin{align}
	\begin{split}
	\begin{bmatrix}
	\textbf{p}^{\phi}_L(\ell-1,\cdot,q_{0})\\
	\vdots\\
	\textbf{p}^{\phi}_L(\ell-1,\cdot,q_{K})\\	
	\end{bmatrix}=
	\begin{bmatrix}
	\textbf{C}^{\rm{p}}_{0,0}  & \dots & \textbf{C}^{\rm{p}}_{0,K}\\
	\vdots & \vdots & \vdots \\
	\textbf{C}^{\rm{p}}_{K,0} & \dots & \textbf{C}^{\rm{p}}_{K,K}\\
	\end{bmatrix}
	\begin{bmatrix}
	\textbf{p}^{\phi}_L(\ell,\cdot,q_{0})\\
	\vdots\\
	\textbf{p}^{\phi}_L(\ell,\cdot,q_{K})\\
	\end{bmatrix}	
	\end{split}
	\label{recursive3}
	\end{align}
	where $\textbf{C}^{\rm{p}}_{j,k}(i_1, i_2)=P^{\phi}_{\rm{p}}((s^{i_1},q_j),(s^{i_2},q_k))$, $\textbf{p}^{\phi}_L(\ell,\cdot,q_{k}):=[p^{\phi}_L(\ell,s^1,q_{k}), p^{\phi}_L(\ell,s^2,q_{k}), \dots, p^{\phi}_L(\ell,s^H,q_{k})]^T$.
	
	Finally, we have $\mathbb{P}_{\mathcal{B}_L,\phi}=\sum\limits_{s^i\in S_0}p_{\rm{int}}(s^i)p^{\phi}_L(0,s^i,q_{0})$, where $p_{\rm{int}}(s^i)$ is the probability of the initial state being $s^i$.
	
	For a syntactically safe pLTL formula $\phi$, we use $p^{\lnot\phi}_L(\ell;s^i,q_{k})$ to denote the probability of extending a path of time length $\ell$ to time length $L$, with the state of the product automaton $\mathcal{M}^{\lnot\phi}_L$ (of $\mathcal{M}$ and $\mathcal{A}^{\lnot\phi}_L$) at time instant $\ell$ being the state $(s^i, q_{k})$, such that the extended path of time length $L$ strongly violates $\phi$ (i.e. strongly satisfies $\lnot\phi$).   
Then we have the following:
\[
\begin{split}
p^{\lnot\phi}_L(L;s^i,q_{j})=                                                            
\begin{cases}
1  ~~~~~~~~~~~~~~~\mbox{if}~q_j\in Acc^{\lnot\phi};\\ 
0 ~~~~~~~~~~~~~~~~\mbox{otherwise}.
\end{cases}
\end{split}
\]
We can compute $p^{\lnot\phi}_L(\ell;s^i,q_{k})$ recursively using (\ref{recursive3}) by replacing each $\phi$ with $\lnot\phi$. Finally, we have
$\mathbb{P}_{\mathcal{B}_L,\phi}=\sum\limits_{s^i\in S_0}p_{\rm{int}}(s^i)(1-p^{\lnot\phi}_L(0;s^i,q_{0}))$. 

With $\mathbb{P}_{\mathcal{B}_L,\phi}$, we can calculate $\mathcal{I}(\mathcal{F}_L,\mathcal{\bar{F}}^{\phi}_L)$ according to Proposition \ref{exp}.

\bibliographystyle{IEEEtran}
\bibliography{zherefclean_submit}

\end{document}